\documentstyle[12pt]{article}
\newcommand{\dfrac}{\displaystyle \frac}
\newcommand{\Dsla}{D\hspace{-9.25pt} /}
\def\ao{{}\kern-.10em\hbox{``}}
\begin{document}
\large
\bibliographystyle{plain}
\begin{titlepage}
\hfill \begin{tabular}{l} HEPHY-PUB 597/94\\ UWThPh-1994-04\\ March 1994
\end{tabular}\\[4cm]
\begin{center}
{\Large \bf A PARTICULAR CLASS OF GUTS WITH VANISHING ONE-LOOP BETA
FUNCTIONS}\\
\vspace{1.5cm}
{\Large \bf Wolfgang LUCHA}\\[.5cm]
{\large Institut f\"ur Hochenergiephysik\\
\"Osterreichische Akademie der Wissenschaften\\
Nikolsdorfergasse 18, A-1050 Wien, Austria}\\[1cm]
{\Large \bf Franz F. SCH\"OBERL}\\[.5cm]
{\large Institut f\"ur Theoretische Physik\\
Universit\"at Wien\\
Boltzmanngasse 5, A-1090 Wien, Austria}\\[3cm]
{\bf Abstract}
\end{center}
\normalsize

\noindent
By explicit solution of the one-loop finiteness conditions for gauge and
quartic scalar-boson self-interaction coupling constants, a particular class
of grand unified theories with vanishing Yukawa couplings as well as
vanishing one-loop renormalization-group beta functions is constructed.
\end{titlepage}

\section{Introduction}

The most important feature of supersymmetry, which attracted in the past a
lot of interest, is its particularly far-reaching ability of softening the
high-energy behaviour of quantum field theories, which culminates in the
possibility of constructing a certain class of perturbatively finite quantum
field theories, namely, the well-known $N = 2$ supersymmetric theories
satisfying a single and only one-loop \ao finiteness condition."
\footnote{\normalsize\ In general, any additional symmetry of some quantum
field theory diminishes the number of uncorrelated ultraviolet divergences.
Supersymmetry plays a special r\^{o}le in reducing this number eventually to
zero.} (For some review of these developments see, for instance, Ref.
\cite{shelter83}.) It soon became clear that the actual r\^{o}le of
supersymmetry for finiteness may only be revealed by subjecting the most
general renormalizable quantum field theory to the requirement of finiteness
\cite{lucha86a,lucha86b,lucha87a}, in order to see whether the constraints of
finiteness admit non-supersymmetric solutions too. (One has to bear in mind,
however, that every eventual conclusion may depend on the chosen
renormalization scheme \cite{lucha87b} since only regularization methods
which involve dimensional regularization parameters are able to take into
account also quadratic divergences.)

Within this context, in Ref. \cite{shapiro93} two classes of
non-supersymmetric one-loop finite grand unified models with, in
group-theoretical respect, particularly simple matter content have been
introduced; the strengths of gauge, Yukawa, and scalar-boson self-couplings
are determined by demanding the one-loop contributions to their
renormalization-group beta functions to vanish. Nevertheless, from the
investigation of their eventual quadratic divergences in Ref. \cite{lucha93a}
it may be deduced that the vector-boson masses of both classes of models as
well as the scalar-boson masses of the whole class of simpler models and of
all explicitly constructed representatives of the class of more sophisticated
models receive already at one-loop level quadratically divergent
contributions. In Ref. \cite{lucha93b}, we succeeded in constructing the
complete class of simpler models while for the class of more complicated
models, because of the complexity of the corresponding finiteness conditions
even at one-loop level, we were only able to derive its most important
general features. Here, we construct explicitly a particular subset of the
class of general models, namely, that one which is characterized by vanishing
Yukawa couplings (Sect.~\ref{sec:model}), by matching their behaviour for
large gauge groups (Sect.~\ref{sec:asympbehav}) with the solutions
\cite{lucha93b} known for small gauge groups (Sect.~\ref{sec:sumconc}).

\section{The \ao Non-Yukawa" Model}\label{sec:model}

The model---or, more precisely, class of models---under consideration may be
deduced from the so-called \ao general" model proposed in Ref.
\cite{shapiro93} and discussed in Refs. \cite{lucha93a,lucha93b} by imposing
on the latter the additional requirement of the vanishing of all Yukawa
interactions. Consequently, it constitutes a completely massless gauge theory
based on the special unitary group $\mbox{SU($N$)}$ as gauge group. The
generators of $\mbox{SU($N$)}$ in its fundamental representation will be
denoted by $T^a$, $a = 1,2,\dots,N^2 - 1$. Their normalization is determined
by the freely chosen value of their second-order Dynkin index $T_{\rm f}$,
defined according to $T_{\rm f}\,\delta_{ab} := \mbox{Tr}(T^a\,T^b)$. These
generators satisfy the commutation relations $[T^a,T^b] = i\,f_{abc}\,T^c$,
with $f_{abc}$ the completely antisymmetric structure constants of SU($N$).
The particle content of this model comprises
\begin{itemize}
\item (real) gauge vector bosons $V_\mu^a$, demanded by gauge invariance and
entering in both the field strength $F_{\mu\nu}^a \equiv \partial_\mu V_\nu^a
- \partial_\nu V_\mu^a + g\,f_{abc}\,V_\mu^b\,V_\nu^c$ and the covariant
derivative $D_\mu \equiv \partial_\mu - i\,g\,V_\mu^a\,T^a$,
\item $m$ sets of Dirac fermions $\Psi_{(k)}$, $k = 0,1,2,\dots ,m$, and
\item real scalar bosons $\Phi$,
\end{itemize}
all of them transforming according to the adjoint representation of the gauge
group, as well as
\begin{itemize}
\item $n$ sets of Dirac fermions $\psi_{(k)}$, $k = 0,1,2,\dots ,n$, and
\item complex scalar bosons $\varphi$,
\end{itemize}
all of them transforming according to the fundamental representation of the
gauge group. The Lagrangian defining this model reads
\begin{eqnarray}
{\cal L}
&=& - \dfrac{1}{4} \, F_{\mu\nu}^a \, F^{\mu\nu}_a
+ i \sum_{k = 0}^m \bar\Psi_{(k)} \, \Dsla \: \, \Psi_{(k)}
+ i \sum_{k = 0}^n \bar\psi_{(k)} \, \Dsla \: \, \psi_{(k)} \nonumber\\
&+& \dfrac{1}{2} \left( D_\mu \Phi \right)^T D^\mu \Phi
+ \left( D_\mu \varphi \right)^\dagger D^\mu \varphi
- \dfrac{\lambda_1}{8} \left( \Phi^T \Phi \right)^2
- \dfrac{\lambda_2}{8} \left( \Phi^a \, d_{abc} \, \Phi^b \right)^2
\nonumber\\
&-& \dfrac{\lambda_3}{2} \left( \Phi^T \Phi \right) \left( \varphi^\dagger
\varphi \right)
- \dfrac{\lambda_4}{2} \left( \Phi^a \, d_{abc} \, \Phi^b \right) \left(
\varphi^\dagger \, T^c \, \varphi \right)
- \dfrac{\lambda_5}{2} \left( \varphi^\dagger \varphi \right)^2 \quad ,
\nonumber
\end{eqnarray}
where $d_{abc}$ are the completely symmetric constants
$$
d_{abc} \equiv
\dfrac{\mbox{Tr}\left(\left\{T^a,T^b\right\}T^c\right)}{T_{\rm f}} \quad .
$$
The absence of Yukawa couplings at higher orders of the perturbative loop
expansion, obligatory for maintenance of renormalizability of the theory, is
guaranteed by the invariance of the above Lagrangian under the reflections
$\Phi \to - \Phi$ and $\varphi \to - \varphi$.

Once the group-theoretic affairs have been settled, the only physical
parameters of this model are the gauge coupling constant, $g$, and the five
scalar-boson self-interaction coupling constants
$\lambda_1,\lambda_2,\dots,\lambda_5$. The (relative) magnitudes of these
dimensionless coupling constants will be determined from the requirement of
vanishing one-loop contributions to their renormalization-group beta
functions.

For the gauge coupling constant, finiteness at the one-loop level is ensured
if the \ao group parameter" $N$ and the fermion multiplicities $m$ and $n$
are related by
$$
21 \, N - 4 \, (2 \, m \, N + n) = 1 \quad .
$$
This constraint forces the group parameter $N$ to take one of the values $N =
4\,\ell + 1$ with $\ell = 1,2,\dots$, that is, one of the values $N =
5,9,13,\dots$, and, from the (necessary) positivity of the multiplicity $n$,
i.~e., $n \ge 0$, implies for the multiplicity $m$ the upper bound
$$
m \le \dfrac{21 \, N - 1}{8 \, N} < 3
\quad \mbox{for arbitrary $N > 0$} \quad ,
$$
which restricts $m$ to any of the three integers $m = 0,1,2$ and, for any
particular choice of $m$, fixes $n$ to just that value which allows to fulfil
this constraint.

\section{Asymptotic Behaviour for Large Gauge Groups}\label{sec:asympbehav}

In order to find out by purely algebraic means the general structure to be
expected for the solutions of the set of one-loop finiteness conditions for
the quartic scalar-boson self-couplings
$\lambda_1,\lambda_2,\lambda_3,\lambda_4,\lambda_5$, we consider first the
large-$N$ limit of these finiteness conditions.

In the course of the analysis of Ref. \cite{lucha93b} it has proven very
convenient to perform the investigation of the conditions for the vanishing
of the one-loop contributions \cite{shapiro93} to the renormalization-group
beta functions of the quartic scalar-boson self-couplings $\lambda_i$, $i =
1,2,\dots,5$, in terms of the five real and non-negative \ao
self-interaction-type" variables
$$
y_i \equiv \dfrac{\lambda_i}{g^2} \ge 0 \quad , \qquad i = 1,2,\dots ,5
\quad .
$$
The asymptotic form the set of one-loop finiteness conditions for the quartic
scalar-boson self-interaction coupling constants assumes in the limit $N \to
\infty$ may immediately be read off from its general form given by Eqs. (43)
to (47) of Ref. \cite{lucha93b}:
\begin{eqnarray}
&&N^2 \, y_1^2 + 24 - 12 \, N \, y_1 + 4 \, N \, y_1 \, y_2 + 8 \, y_2^2
+ 2 \, N \, y_3^2 = 0 \quad ,
\nonumber \\[1ex]
&&4 \, N \, y_2^2 - 12 \, N \, y_2 + 3 \, N + 12 \, y_1 \, y_2 + y_4^2 = 0
\quad ,
\nonumber \\[1ex]
&&4 \, y_3^2 - 9 \, N \, y_3 + 6 + N^2 \, y_1 \, y_3 + 2 \, N \, y_2 \, y_3
+ 2 \, N \, y_3 \, y_5 + 2 \, y_4^2 = 0 \quad ,
\nonumber \\[1ex]
&&N \, y_4^2 - 9 \, N \, y_4 + 3 \, N + 2 \, y_1 \, y_4 + 2 \, N \, y_2 \, y_4
+ 2 \, y_4 \, y_5 + 8 \, y_3 \, y_4 = 0 \quad ,
\nonumber \\[1ex]
&&4 \, y_5^2 - 12 \, y_5 + 2 \, N \, y_3^2 + y_4^2 + 3 = 0 \quad .
\nonumber
\end{eqnarray}

Each of the (even general) one-loop finiteness conditions for the five
self-interaction-type variables $y_1,y_2,y_3,y_4,y_5$ involves a single
negative term linear in precisely one of the variables $y_1$, $y_2$, $y_3$,
$y_4$, or $y_5$, which is therefore responsible for counterbalancing all
positive terms. In Ref. \cite{lucha93b} it has been demonstrated that, as a
consequence of the presence of these single negative terms, the variables
$y_1,y_2,\dots ,y_5$ are, for $N \ge 5$, bounded by
\begin{eqnarray}
\dfrac{2}{N} < &y_1& < \dfrac{9}{N} \quad ,
\nonumber \\[1ex]
\dfrac{1}{4} < &y_2& < \dfrac{25}{7} \quad ,
\nonumber \\[1ex]
\dfrac{2}{3 \, N} < &y_3& < \dfrac{9 \, N}{8} \quad ,
\nonumber \\[1ex]
\dfrac{1}{3} < &y_4& < \dfrac{222}{13} \quad ,
\nonumber \\[1ex]
\dfrac{1}{4} < &y_5& < 3 \quad .
\nonumber
\end{eqnarray}

Consider a generic positive variable $y$ depending on some parameter
$\lambda$, that is, $y = y(\lambda) > 0$ for all $1 < \lambda < \infty$. Let
this variable be bounded from below and above by two bounds which scale like
some powers $\alpha$ and $\beta$ of the parameter $\lambda$, respectively. In
other words, let the variable $y$ satisfy a chain of inequalities of the form
$$
0 < a \, \lambda^\alpha < y(\lambda) < b \, \lambda^\beta < \infty \qquad
\mbox{for all} \quad 1 < \lambda < \infty \quad ,
$$
with some positive constants $a$ and $b$. In view of that, assume for this
variable $y$ a similar power-law behaviour,\footnote{\normalsize\ This
assumption precludes, of course, any oscillatory dependence of $y$ on
$\lambda$.} with some exponent $\gamma$ and some positive constant $k$:
$$
y(\lambda) = k \, \lambda^\gamma \quad , \qquad 0 < k < \infty \quad .
$$
Then the exponent $\gamma$ is necessarily confined to the range spanned by
the two powers $\alpha$ and $\beta$:
$$
\alpha \le \gamma \le \beta \quad .
$$
Consequently, if, in particular, the two powers $\alpha$ and $\beta$ are
equal, i.~e., for $\alpha = \beta$, the exponent $\gamma$ has to share this
common value,
$$
\alpha = \gamma = \beta \quad ,
$$
and, in addition, the constant $k$ is bounded by the constants $a$ and $b$,
$$
a < k < b \quad .
$$

Adhering to these lines of argumentation, we assume that our five
self-interaction-type variables $y_1,y_2,y_3,y_4,y_5$ satisfy the above
bounds by exhibiting, for sufficiently large values of the group parameter
$N$, a power-law behaviour parametrized by five constants,
$k_1,k_2,k_3,k_4,k_5$, and the only exponent not fixed by the bounds on these
variables, $\gamma$:
\begin{eqnarray}
y_1 &=& \dfrac{k_1}{N} \quad , \qquad 2 < k_1 < 9 \quad ,
\nonumber \\[1ex]
y_2 &=& k_2 \quad , \qquad \dfrac{1}{4} < k_2 < \dfrac{25}{7} \quad ,
\nonumber \\[1ex]
y_3 &=& k_3 \, N^\gamma \quad , \qquad - 1 \le \gamma \le + 1 \quad ,
\nonumber \\[1ex]
y_4 &=& k_4 \quad , \qquad \dfrac{1}{3} < k_4 < \dfrac{222}{13} \quad ,
\nonumber \\[1ex]
y_5 &=& k_5 \quad , \qquad \dfrac{1}{4} < k_5 < 3 \quad .
\nonumber
\end{eqnarray}
Inserting this ansatz into the above self-coupling finiteness conditions and
recalling again the limit $N \to \infty$, the set of equations which serves
to pin down the asymptotic behaviour of the five self-interaction-type
variables $y_1,y_2,y_3,y_4,y_5$ becomes
\begin{eqnarray}
&&k_1^2 + 24 - 12 \, k_1 + 4 \, k_1 \, k_2 + 8 \, k_2^2
+ 2 \, N^{1 + 2 \, \gamma} \, k_3^2 = 0 \quad ,
\label{eq:fc(1)k}\\[1ex]
&&4 \, k_2^2 - 12 \, k_2 + 3 = 0 \quad ,
\label{eq:fc(2)k}\\[1ex]
&&4 \, N^{2 \, \gamma} \, k_3^2 + N^{1 + \gamma}\, k_3 \, (k_1 + 2 \, k_2
+ 2 \, k_5 - 9) + 6 + 2 \, k_4^2 = 0 \quad ,
\label{eq:fc(3)k}\\[1ex]
&&k_4^2 - 9 \, k_4 + 3 + 2 \, k_2 \, k_4 + 8 \, N^{\gamma - 1} \, k_3 \, k_4
= 0 \quad ,
\label{eq:fc(4)k}\\[1ex]
&&4 \, k_5^2 - 12 \, k_5 + 2 \, N^{1 + 2 \, \gamma} \, k_3^2 + k_4^2 + 3 = 0
\quad .
\label{eq:fc(5)k}
\end{eqnarray}

Now, the group parameter, $N$, enters into each of Eqs. (\ref{eq:fc(1)k}) and
(\ref{eq:fc(5)k}) only via one and the same single term, and that with the
power $1 + 2\,\gamma$. Consequently, for non-zero $k_3$ as required by the
above lower bound on the variable $y_3$, a solution to either of Eqs.
(\ref{eq:fc(1)k}) and (\ref{eq:fc(5)k}) can only exist if the exponent
$\gamma$ is bounded from above by $\gamma \le - \frac{1}{2}$. Accordingly,
the conceivable range of the exponent $\gamma$ is narrowed down to $- 1 \le
\gamma \le - \frac{1}{2}$. For $\gamma$ within this range, the set of
equations (\ref{eq:fc(1)k}) to (\ref{eq:fc(5)k}) simplifies, once more
because of the limit $N \to \infty$ implicitly understood, to
\begin{eqnarray}
&&k_1^2 + 24 - 12 \, k_1 + 4 \, k_1 \, k_2 + 8 \, k_2^2
+ 2 \, N^{1 + 2 \, \gamma} \, k_3^2 = 0 \quad ,
\label{eq:fc(1)k-1/2}\\[1ex]
&&4 \, k_2^2 - 12 \, k_2 + 3 = 0
\quad ,
\label{eq:fc(2)k-1/2}\\[1ex]
&&N^{1 + \gamma}\, k_3 \, (k_1 + 2 \, k_2 + 2 \, k_5 - 9) + 6 + 2 \, k_4^2 = 0
\quad ,
\label{eq:fc(3)k-1/2}\\[1ex]
&&k_4^2 - 9 \, k_4 + 3 + 2 \, k_2 \, k_4 = 0 \quad ,
\label{eq:fc(4)k-1/2}\\[1ex]
&&4 \, k_5^2 - 12 \, k_5 + 2 \, N^{1 + 2 \, \gamma} \, k_3^2 + k_4^2 + 3 = 0
\quad .
\label{eq:fc(5)k-1/2}
\end{eqnarray}
The detailed investigation of the case $- 1 < \gamma \le - \frac{1}{2}$,
however, indicates that for $\gamma > - 1$ there exists no solution at all to
the set of equations (\ref{eq:fc(1)k-1/2}) to (\ref{eq:fc(5)k-1/2}). The
corresponding reasoning is the following. On the one hand, Eq.
(\ref{eq:fc(2)k-1/2}) may immediately be solved for $k_2$. After that, with
both possible solutions for $k_2$ at hand, Eq. (\ref{eq:fc(4)k-1/2}) may be
solved for $k_4$. On the other hand, for $\gamma > - 1$ Eq.
(\ref{eq:fc(3)k-1/2}) reduces to $k_1 + 2 \, k_2 + 2 \, k_5 - 9 = 0$. Eqs.
(\ref{eq:fc(1)k-1/2}), (\ref{eq:fc(3)k-1/2}), and (\ref{eq:fc(5)k-1/2}) may
then be combined to yield an expression for $k_4$, namely, $k_4^2 = 4 \,
k_2^2 + 24 \, k_2 - 6$, which is, already by the mere number of possible
solutions, in clear conflict with Eq. (\ref{eq:fc(4)k-1/2}). Consequently,
the up to now indeterminate value of the exponent $\gamma$ is unambiguously
fixed to $\gamma = - 1$. In summary, the large-$N$ behaviour of our
self-interaction-type variables $y_1,y_2,\dots,y_5$ is described by
$$
y_1 = \dfrac{k_1}{N} \quad , \quad y_2 = k_2 \quad , \quad
y_3 = \dfrac{k_3}{N} \quad , \quad y_4 = k_4 \quad , \quad y_5 = k_5 \quad ,
$$
where the five constants $k_1,k_2,\dots,k_5$ are to be extracted from the set
of equations
\begin{eqnarray}
&&k_1^2 + 24 - 12 \, k_1 + 4 \, k_1 \, k_2 + 8 \, k_2^2 = 0 \quad ,
\label{eq:fc(1)k-1}\\[1ex]
&&4 \, k_2^2 - 12 \, k_2 + 3 = 0 \quad ,
\label{eq:fc(2)k-1}\\[1ex]
&&k_3 \, (k_1 + 2 \, k_2 + 2 \, k_5 - 9) + 6 + 2 \, k_4^2 = 0 \quad ,
\label{eq:fc(3)k-1}\\[1ex]
&&k_4^2 - 9 \, k_4 + 3 + 2 \, k_2 \, k_4 = 0 \quad ,
\label{eq:fc(4)k-1}\\[1ex]
&&4 \, k_5^2 - 12 \, k_5 + k_4^2 + 3 = 0 \quad .
\label{eq:fc(5)k-1}
\end{eqnarray}

In this final form the set of self-coupling finiteness conditions valid in
the limit $N \to \infty$ contains, of course, no longer any reference to the
group parameter $N$. Therefore, owing to its specific internal structure, it
is straightforward to derive step by step its complete set of solutions.
First of all, Eq. (\ref{eq:fc(2)k-1}) is (and has been already from the very
beginning) a quadratic equation for $k_2$ only and may hence immediately be
solved for $k_2$. Then, for a given $k_2$, Eq. (\ref{eq:fc(1)k-1}) reduces to
a quadratic equation for $k_1$ only and may thus be solved for $k_1$. Reality
of $k_1$ eliminates one of the two solutions of Eq. (\ref{eq:fc(2)k-1}) for
$k_2$, leaving a unique solution for $k_2$. Simultaneously, again for a given
$k_2$, Eq. (\ref{eq:fc(4)k-1}) reduces to a quadratic equation for $k_4$ only
and may hence be solved for $k_4$. After that, for a given $k_4$, Eq.
(\ref{eq:fc(5)k-1}) reduces to a quadratic equation for $k_5$ only and may
thus be solved for $k_5$. Reality of $k_5$ eliminates one of the two
solutions of Eq. (\ref{eq:fc(4)k-1}) for $k_4$, leaving a unique solution for
$k_4$. Finally, for a given $k_1$, $k_2$, $k_4$, and $k_5$, Eq.
(\ref{eq:fc(3)k-1}) reduces to a linear equation for $k_3$ only, which
unambiguously entails $k_3$. Positivity of $k_3$ eliminates one of the two
solutions of Eq. (\ref{eq:fc(5)k-1}) for $k_5$, leaving a unique solution for
$k_5$. All together, we end up with the following expressions for the
constants $k_1,k_2,\dots,k_5$:
\begin{eqnarray}
k_1 &=& 3 + \sqrt{6} \pm \sqrt{3\left(6\sqrt{6} - 13\right)}
= \left\{\begin{tabular}{r}7.7057\dots\\ 3.1932\dots\end{tabular}\right.
\quad , \nonumber\\[1ex]
k_2 &=& \dfrac{1}{2} \left(3 - \sqrt{6}\right)
= 0.2752\dots \quad , \nonumber\\[1ex]
k_3 &=& 2\sqrt{3} \, \dfrac{7 + 2\sqrt{6} - \left(\sqrt{6} + 1\right)
\sqrt{5 + 2\sqrt{6}}}{\sqrt{\left(\sqrt{6} + 1\right)
\sqrt{5 + 2\sqrt{6}} - 4 - 2\sqrt{6}} \mp \sqrt{6\sqrt{6} - 13}} \nonumber\\
&=& \left\{\begin{tabular}{r}38.0607\dots\\ 1.3417\dots\end{tabular}\right.
\quad , \nonumber\\[1ex]
k_4 &=&
\sqrt{\dfrac{3}{2}} \left(\sqrt{6} + 1 - \sqrt{5 + 2\sqrt{6}} \, \right)
= 0.3713\dots \quad , \nonumber\\[1ex]
k_5 &=& \dfrac{1}{2} \left(3 - \sqrt{3\left[\left(\sqrt{6} + 1\right)
\sqrt{5 + 2\sqrt{6}} - 4 - 2\sqrt{6}\right]} \, \right) \nonumber\\
&=& 0.2894\dots \quad . \nonumber
\end{eqnarray}
The asymptotic set of self-interaction finiteness conditions, Eqs.
(\ref{eq:fc(1)k-1}) to (\ref{eq:fc(5)k-1}), thus possesses precisely two sets
of solutions with all members real and non-negative, and respecting, of
course, their above bounds.

\section{Summary and Conclusions}\label{sec:sumconc}

The present note has been devoted to the complete construction of a
particular class of grand unified theories; the members of this class are
characterized by the two requirements of vanishing Yukawa couplings and
vanishing one-loop beta functions for the gauge and scalar-boson
self-interaction coupling constants. Quite generally, one-loop finiteness of
the gauge coupling constant may be trivially satisfied by a suitable choice
of the matter content of the theory under consideration. Hence, in order to
get an idea of the spectrum of solutions to be expected for the whole set of
finiteness conditions, we investigated in the preceding section the relations
entailed by one-loop finiteness of the scalar-boson self-interaction coupling
constants in the limit of infinitely large gauge groups. We were able to
identify precisely two distinct sets of solutions for the scalar-boson
self-couplings.

Now, the actual object of our interest is, of course, the spectrum of
solutions for gauge groups of finite size. The respective set of one-loop
finiteness conditions for the self-interaction-type variables
$y_1,y_2,\dots,y_5$ is given by Eqs. (43) through (47) of Ref.
\cite{lucha93b}. Furthermore, according to Ref. \cite{lucha93b}, the smallest
possible gauge group allowing for a solution of these general finiteness
conditions is SU(9). In Ref. \cite{lucha93b}, two and only two sets of
solutions for the self-interaction-type variables $y_1,y_2,\dots,y_5$ have
been found for the gauge group SU(9); we reproduce them here, for the sake of
completeness, in Table \ref{tab:nonyuk-N=9}.
{\normalsize
\begin{table}[h]
\begin{center}
\caption[]{Numerical solutions for the five self-interaction-type variables
$y_1,y_2,y_3,y_4,y_5$ of the \ao non-Yukawa" model based on the gauge group
SU($9$) (taken from Ref. \cite{lucha93b}) as well as the corresponding
quantities $N\,y_1$ and $N\,y_3$.}\label{tab:nonyuk-N=9}
\vspace{0.5cm}
\begin{tabular}{|c|l|l|}
\hline
&&\\[-1ex]
\multicolumn{1}{|c|}{Variable}&\multicolumn{1}{c|}{Solution I}&
\multicolumn{1}{c|}{Solution II}\\
&&\\[-1.5ex]
\hline
&&\\[-1.5ex]
$\quad y_1\quad$&$\quad$0.4017\dots$\quad$&$\quad$0.5054\dots$\quad$\\
$\quad y_2\quad$&$\quad$0.2864\dots$\quad$&$\quad$0.2907\dots$\quad$\\
$\quad y_3\quad$&$\quad$0.1862\dots$\quad$&$\quad$0.3676\dots$\quad$\\
$\quad y_4\quad$&$\quad$0.3860\dots$\quad$&$\quad$0.4008\dots$\quad$\\
$\quad y_5\quad$&$\quad$0.4167\dots$\quad$&$\quad$0.7812\dots$\quad$\\
&&\\[-1.5ex]
\hline
&&\\[-1.5ex]
$\quad N\,y_1\quad$&$\quad$3.6161\dots$\quad$&$\quad$4.5491\dots$\quad$\\
$\quad N\,y_3\quad$&$\quad$1.6762\dots$\quad$&$\quad$3.3092\dots$\quad$\\
[1.5ex]
\hline
\end{tabular}
\end{center}
\end{table}}

Our findings at the above two opposite extremes of possible values of the
group parameter $N$, that is, $N = 9$ and $N \to \infty$, may easily be
reconciled by tracing numerically, for $N$ covering its allowed range, the
variation of the solutions of the general one-loop finiteness conditions.
Qualitatively, the following behaviour of our five self-interaction-type
variables $y_1,y_2,y_3,y_4,y_5$ with increasing group parameter $N$ emerges:
\begin{itemize}
\item For the quantity $N\,y_1$, the larger solution increases monotonously
from the value $N\,y_1 = 4.5491\dots$ at $N = 9$ towards its asymptotic value
$k_1 = 7.7057\dots$ for the limit $N \to \infty$ whereas the smaller solution
decreases monotonously from the value $N\,y_1 = 3.6161\dots$ at $N = 9$
towards its asymptotic value $k_1 = 3.1932\dots$ for the limit $N \to
\infty$.
\item For the variable $y_2$, both of the solutions decrease monotonously
from their values $y_2 = 0.2907\dots$ and $y_2 = 0.2864\dots$, respectively,
at $N = 9$ towards their common asymptotic value $k_2 = 0.2752\dots$ for the
limit $N \to \infty$.
\item For the quantity $N\,y_3$, the larger solution increases monotonously
from the value $N\,y_3 = 3.3092\dots$ at $N = 9$ towards its asymptotic value
$k_3 = 38.0607\dots$ for the limit $N \to \infty$ whereas the smaller
solution decreases monotonously from the value $N\,y_3 = 1.6762\dots$ at $N =
9$ towards its asymptotic value $k_3 = 1.3417\dots$ for the limit $N \to
\infty$.
\item For the variable $y_4$, both of the solutions decrease monotonously
from their values $y_4 = 0.4008\dots$ and $y_4 = 0.3860\dots$, respectively,
at $N = 9$ towards their common asymptotic value $k_4 = 0.3713\dots$ for the
limit $N \to \infty$.
\item For the variable $y_5$, interestingly, the larger solution first starts
to increase from the value $y_5 = 0.7812\dots$ at $N = 9$ and then, after
passing its maximum value $y_5 = 1.2742\dots$ at $N = 29$, continues to
decrease in order to approach finally the common asymptotic value $k_5 =
0.2894\dots$ for the limit $N \to \infty$ whereas the smaller solution
decreases monotonously from the value $y_5 = 0.4167\dots$ at $N = 9$ towards
this common asymptotic value $k_5 = 0.2894\dots$ for the limit $N \to
\infty$.
\end{itemize}

In summary, by the above findings we are led to conclude that for each gauge
group allowed by one-loop finiteness of the gauge coupling constant there
will exist exactly two \ao non-Yukawa" models of the kind described in Sect.
\ref{sec:model}; the numerical values of the quartic scalar-boson
self-interaction coupling constants (relative to the square of the gauge
coupling constant) are fixed to be precisely those which guarantee the
vanishing of the one-loop contributions to their renormalization-group beta
functions.

Finally, concerning the question of eventual quadratic divergences, within
the present class of models both vector-boson and scalar-boson masses will be
plagued by quadratic divergences already from one-loop level on: On the one
hand, Eqs. (17) and (25) of Ref. \cite{lucha93a} state that for both classes
of models proposed in Ref. \cite{shapiro93} the quadratically divergent
contribution to the one-loop renormalization of the vector-boson mass is
definitely non-vanishing. On the other hand, since the present class of
models is a subset of the class of general models considered in Ref.
\cite{lucha93a}, we infer from Eqs. (18) and (19) of Ref. \cite{lucha93a}
that the quadratically divergent contributions to the one-loop
renormalization of the masses of both kinds of scalar bosons, $\Phi$ and
$\varphi$, must be non-vanishing for the whole class of non-Yukawa models;
the trivial reason for this being the fact that a non-vanishing Yukawa
interaction is a necessary ingredient for compensating the contributions of
gauge and quartic scalar-boson self-interactions to these renormalizations.

\end{document}